\documentclass[fleqn,twoside,onecolumn,nofootinbib,showkeys,12pt]{revtex4}

\usepackage{graphicx}
\usepackage{amssymb}
\usepackage{amsmath}

\begin{document}

\setcounter{page}{1}%

\begin{center}
\centerline{\large \bf New Signals of Quark-Gluon-Hadron Mixed Phase Formation}

\vspace*{11mm}

%% name corrected
{K.A. Bugaev$^{1,*}$, 
V.V.~Sagun$^1$, 
A.I. Ivanytskyi$^1$, 
D.R. Oliinychenko$^{1,2}$,
E.-M. Ilgenfritz$^3$, 
E.G. Nikonov$^4$, 
A.V. Taranenko$^{5}$ and  G.M.~Zinovjev$^1$}

\vspace*{5.5mm}

{\small

$^1${Bogolyubov Institute for Theoretical Physics of the National Academy of Sciences of Ukraine, Metrologichna str. 14$^B$, Kiev 03680, Ukraine}\\

$^2${FIAS, Goethe University,  Ruth-Moufang Str. 1, 60438 Frankfurt upon Main, Germany}\\

$^3${Bogolyubov Laboratory of Theoretical Physics, JINR, 141980 Dubna, Russia}\\

$^4${Laboratory for Information Technologies, JINR, 141980 Dubna, Russia}\\

$^5${National Research Nuclear University ``MEPhI'' (Moscow Engineering Physics
Institute), Kashirskoe Shosse 31, 115409 Moscow, Russia}

}

%$^*${e-mail: bugaev@th.physik.uni-frankfurt.de}

\end{center}

{\bf Abstract. }{Here we present several remarkable irregularities at chemical freeze-out which are found 
using an advanced version of the hadron resonance gas model. The most prominent of them are the sharp peak of the trace anomaly existing at chemical freeze-out at the center of mass energy 4.9 GeV and two sets of highly correlated quasi-plateaus in the collision energy dependence of the entropy per baryon, total pion number per baryon, and thermal pion number per baryon which we found at the center of mass energies 3.8-4.9 GeV and 7.6-10 GeV. The low energy set of quasi-plateaus was predicted a long time ago. On the basis of the generalized shock-adiabat model we demonstrate that the low energy correlated quasi-plateaus give evidence for the anomalous thermodynamic properties inside the quark-gluon-hadron mixed phase. 
It is also  shown that  the trace anomaly sharp peak at chemical freeze-out corresponds to the trace anomaly peak 
 at the boundary between the mixed phase and quark gluon plasma. We argue
that  the high energy correlated quasi-plateaus may correspond  to a second  phase transition and discuss 
its possible origin and location. Besides we suggest two new observables which may serve as clear signals of these 
phase transformations.\\

\noindent
{\bf PACS}
      {25.75.Nq}{ Quark deconfinement, quark-gluon plasma production,  phase transitions}   
     {12.40.Ee}{ Statistical models}
     } % end of PACS codes

\section{New Irregularities vs New Signals of QGP} 

One of the key problems of modern physics of heavy ion collisions
is formulation of physically justified signals of  the quark-gluon-hadron mixed phase formation \cite{NICA}.
 An essential  progress in formulating such signals was achieved recently \cite{Bugaev:SA1,Bugaev:SA2}. 
 Among the new irregularities related to deconfinement  the most spectacular one is  a sudden jump of the pressure $p$ at chemical freeze-out (CFO) in the narrow range of center of mass  collision energies $\sqrt{s_{NN}} = 4.3-4.9$ GeV  \cite{Bugaev:SA1,SFO:13}.
 The observed pressure jump at CFO is so strong, that the effective number of degrees of freedom, $p/T^4$,  
 increases by 70\%, while the collision energy changes by 15\% or  while the CFO temperature $T$  changes by 30\%.
 
\begin{figure}[t]
\centerline{
\includegraphics[height=96mm,width=98mm]{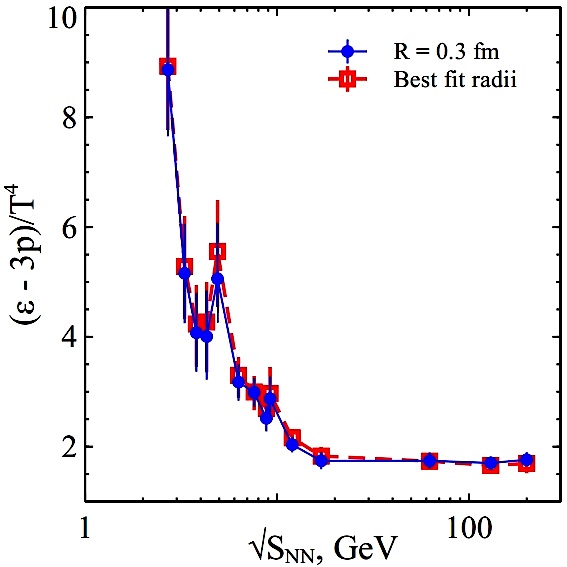}\hspace*{7.7mm}
}
%\hspace*{.5cm}
\centerline{\hspace*{9.7mm}
\includegraphics[height=94mm,width=122mm]{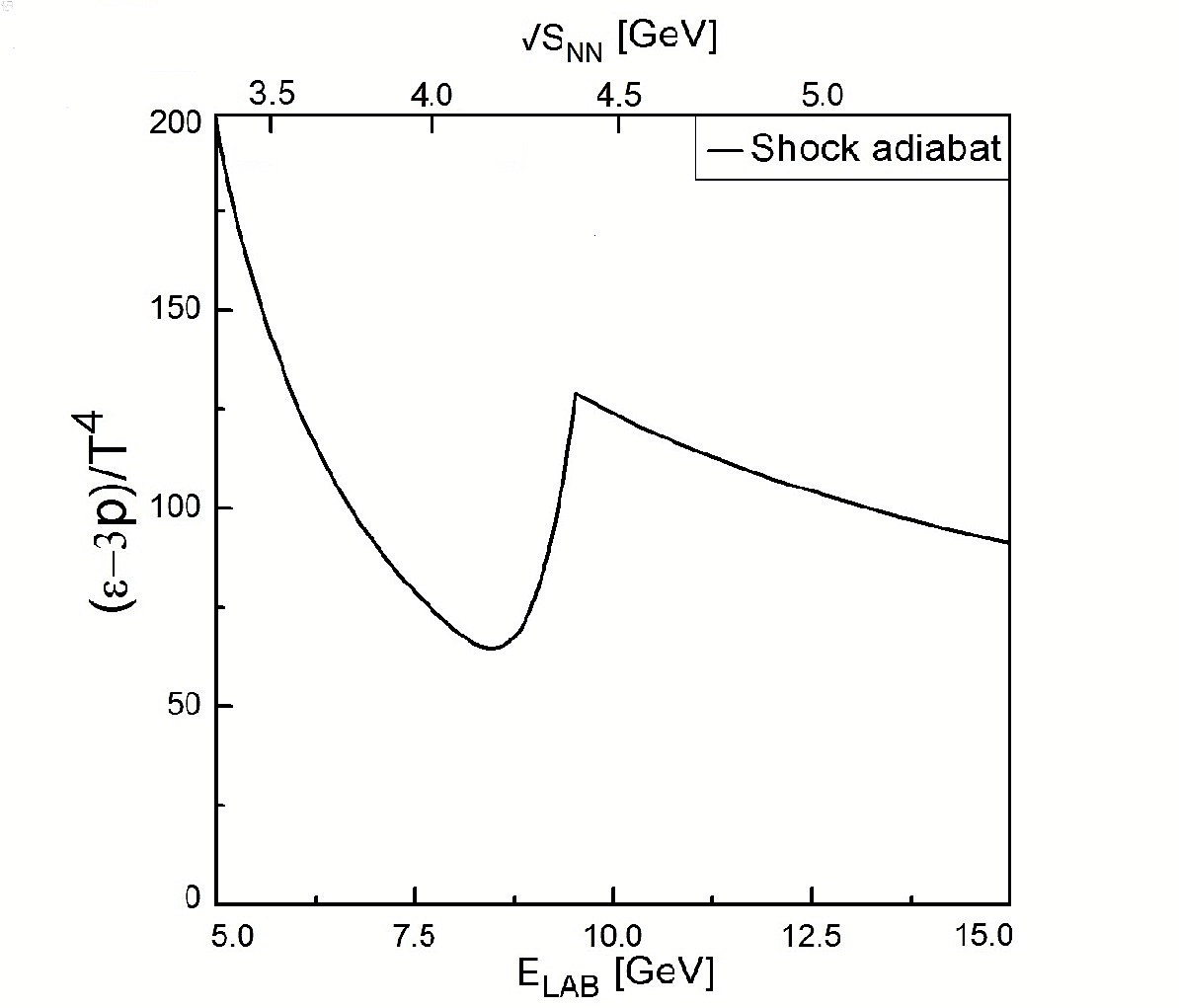}\hspace*{7.7mm} 
}  
 \caption{Trace anomaly as function of collision energy at CFO (upper panel) \cite{Bugaev:SA2} and along the shock  adiabat \cite{Bugaev:SA1} (lower panel).  The details on the set of best fit hard-core radii shown here are given in \cite{SFO:13}.
}
  \label{fig1}
\end{figure}

The other remarkable irregularity found at CFO is the sharp peak of the trace anomaly $\delta = 
\frac{\varepsilon - 3p}{T^4}$ (here $\varepsilon$ is energy density) at $\sqrt{s_{NN}} = 4.9$ GeV  \cite{Bugaev:SA2}
(see the upper panel of Fig.\ \ref{fig1}). 
One could,  of course, conservatively say that
this peak is produced by only one data point and that its significance
should therefore be considered with some caution. However, one can see that the error bars at  
$\sqrt{s_{NN}} = 4.9$ GeV and $\sqrt{s_{NN}} = 6.3$ GeV do not overlap. Moreover,
in \cite{Bugaev:SA2} we found the one to one correspondence between this trace anomaly peak at CFO and 
the trace anomaly peak along the shock adibat (see the lower  panel of Fig.\ \ref{fig1}) which appears at the collision energy corresponding exactly to the boundary between the quark gluon plasma (QGP) and quark-gluon-hadron mixed phase. We remind that the shock adiabat model reasonably well   describes the hydrodynamic and thermodynamic  parameters of the initial state formed in the central nucleus-nucleus collisions in the laboratory 
energy range $1$ GeV $ \le E_{lab} \le $ 30 GeV \cite{Satarov:11}. 
Note also  that the exclusive importance of the found trace anomaly peaks becomes clear, if one recalls that 
the inflection point/maximum of the trace anomaly at low baryonic densities  is traditionally used in lattice 
QCD to determine the pseudo-critical temperature of the cross-over 
transition \cite{lQCD}.

%%%
%%%

\begin{figure}[t]
\centerline{
\hspace*{9.9mm}\includegraphics[height=96mm,width=120mm]{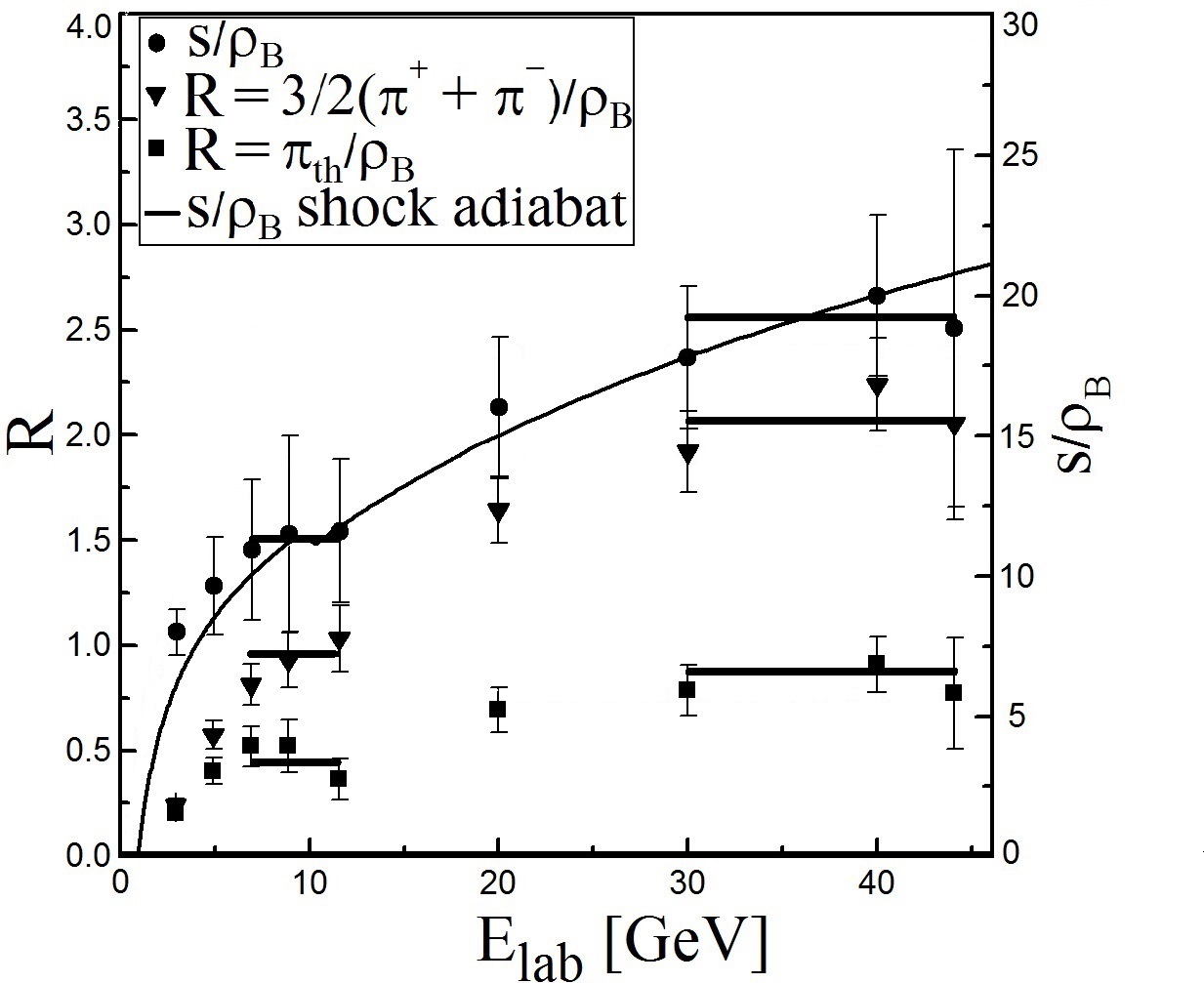} 
}
 \caption{The correlated quasi-plateas at CFO  are taken from \cite{Bugaev:SA1} (see details in the text).}
  \label{fig2a}
\end{figure}

Besides these irregularities the highly correlated quasi-plateaus in the collision energy dependence of the entropy per baryon $s/\rho_B$, total pion number per baryon $\rho_{\pi}^{tot}/\rho_B$, and thermal pion number per baryon $\rho_{\pi}^{th}/\rho_B$ at laboratory energies 6.9--11.6 GeV (i.e. $\sqrt{s_{NN}} = 3.8-4.9$ GeV) were also found in \cite{Bugaev:SA1,Bugaev:SA2}.  As one can see from the  Fig.\ \ref{fig2a} a clear plateau is only seen in the thermal pion number per baryon 
while other quantities exhibit quasi-plateaus. Nevertheless,  these quasi-plateaus  are important, since their strong 
correlation with the plateau  in  
$\rho_{\pi}^{th}/\rho_B$ 
allowed one  to determine their common width in  the laboratory energy \cite{Bugaev:SA1,Bugaev:SA2}.
It is necessary to stress that, in contrast to the most of other CFO irregularities discussed in the modern literature as possible signals of 
QGP formation, these  quasi-plateaus were predicted a long time ago \cite{KAB:89a,KAB:90,KAB:91} on a solid ground of relativistic hydrodynamics. As it was shown in \cite{KAB:89a,KAB:90,KAB:91} and was further argued in \cite{Bugaev:SA1,Bugaev:SA2} the generic reason for an appearance of such quasi-plateaus is in an existence of anomalous thermodynamic properties in the quark-gluon-hadron mixed phase. 
Usually, in pure gaseous or liquid phases the 
interaction between the constituents at short distances is repulsive and, hence, at high  
densities the adiabatic compressibility of matter 
%%%$- \left( \frac{\partial X}{\partial p}\right)_{s/\rho_B}$ 
decreases for  increasing pressure. Hence the pure phases have the normal thermodynamic properties.
In the mixed phase, however,  
there appears another possibility to compress matter: by converting the less dense phase 
into the more dense one.  As it was found for several equations of state  \cite{KAB:89a,KAB:90,KAB:91,Bugaev:SA1,Bugaev:SA2} 
%%%
%%%
with a first-order phase transition 
between hadronic gas and QGP, the phase transition  region has 
anomalous thermodynamic properties, i.e.  it corresponds  to an increase of the
compressibility in the mixed phase at higher pressures.

Using the shock adiabat model  \cite{KAB:89a}, it was possible to  successfully describe the collision energy dependence of the entropy per baryon \cite{Bugaev:SA1} found at CFO (see the solid curve in  Fig.\ \ref{fig2a}). This allowed one to determine the
parameters of the QGP equation of state at high baryonic densities \cite{Bugaev:SA1} directly from CFO data. It was  a great surprise to discover that the whole mixed phase which is rather wide in terms of baryonic densities (it ranges from 2 fm$^{-3}$ at hadronic side of mixed phase to about 10 fm$^{-3}$ at QGP side)  is located between two  neighboring collision energies available in experiments, i.e. inside the interval  $\sqrt{s_{NN}} =4.3-4.9$ GeV.  However, such a narrow collision energy range exactly corresponds to the irregularities discussed above and, moreover, it naturally explains the reason of why the mixed phase was not found during more than three decades of thorough searches.  Moreover, on the basis of shock adiabat model it was possible to explain \cite{Bugaev:SA2} the  other
irregularities discussed above. 

In addition to the low energy set of quasi-plateaus  a second set of  quasi-plateaus  was found at  laboratory 
energies 30--45 GeV ($\sqrt{s_{NN}} = 7.6-9.2$~GeV) \cite{Bugaev:SA1,Bugaev:SA2} as one can see from Fig.\ \ref{fig2a}.  Originally, it was difficult to give a unique  interpretation of this high energy set of quasi-plateaus \cite{Bugaev:SA2}, since the (generalized) shock adiabat model is no longer reliable at these collision energies \cite{Satarov:11}.
Fortunately,  entirely new approach \cite{Metaanalisys} based on a thorough comparison of the quality  of experimental data description by the models which do not assume the QGP existence (hadron models) and which do assume the QGP formation (QGP models)  is able to clarify this question.  Thus, quite independently, the meta-analysis performed in 
\cite{Metaanalisys}  led to a conclusion that the mixed phase exists at the same collision energy range which for the first time was found in \cite{Bugaev:SA1,Bugaev:SA2}, i.e. it exists between $\sqrt{s_{NN}} =4. 3$ GeV and $\sqrt{s_{NN}} =4.9$ GeV.
This result not only  validates the entire framework of shock adiabat model used in \cite{Bugaev:SA1,Bugaev:SA2}, but also it independently justifies the jump of the effective number of degrees of freedom $p/T^4$ at CFO  and a sharp peak of the trace anomaly $\delta$ at CFO as reliable signals of  QGP formation.

Also the meta-analysis of the quality of data description \cite{Metaanalisys} predicts that the most probable  collision energy range of the second phase transition is $\sqrt{s_{NN}} =10-13.5$ GeV. In other words, it independently demonstrates that interpretation of   high energy set of correlated quasi-plateaus  as an indicator of phase transition makes sense, but meta-analysis shifts 
this transition to slightly higher collision energies. 
It is presently not
possible to distinguish two possible explanations of this phenomenon.
Its first explanation is that with increasing collision energy the initial states of  thermally equilibrated matter formed in nucleus-nucleus collisions move from hadron gas into the mixed phase, then from the mixed phase to QGP  and then again they return to the same mixed phase, but at higher initial temperature and lower 
baryonic density. This scenario corresponds to the case, if QCD  phase digram has a critical endpoint \cite{Metaanalisys}. An alternative explanation \cite{Metaanalisys} corresponds to QCD phase diagram with the tricritical endpoint. In this case the second phase transition is the second order phase  transition of (partial) chiral symmetry restoration  or  a transition between quarkyonic matter and QGP \cite{QYON}.  It is remarkable that despite the lack of  a single interpretation of the second phase transition at $\sqrt{s_{NN}} =10-13.5$ GeV there are strong arguments \cite{Metaanalisys} that the (tri)critical endpoint of  QCD phase diagram maybe located within this energy range. 
In what follows we further  develop  the hypothesis  of  two phase transitions, give some additional arguments in favor of (tri)critical endpoint location in the  collision energy range specified above and discuss two new  observables, which 
will be a good indicators of the discussed phase transformations. 
\section{HRGM with Multicomponent Repulsion}
The  uncovering of  novel irregularities discussed in preceding section was achieved using the most 
successful version of the hadron resonance gas model (HRGM) developed during recent years  in  \cite{SFO:13,Oliinychenko:12,HRGM:13,Sagun,Sagun2}.
Existence of  local thermal and chemical equilibrium at CFO allows one to employ the HRGM to describe the hadron yields in the grand canonical formulation  \cite{Andronic:05}, i.e. using the temperature $T$, the baryonic $\mu_B$, the strange $\mu_s$ and the isospin third projection $\mu_{I3}$ chemical potentials.  The chemical potential $\mu_s$ is fixed by the condition of zero total strange charge. 
As usual, the deviation of strange charge from the full chemical equilibrium   is  accounted by the parameter $\gamma_s$
\cite{Rafelski}. 
It modifies the  thermal density of hadron sort $j$ $\varphi_j$ according to the rule $\varphi_j\rightarrow\gamma_s^{S_j}\varphi_j$,
where $S_j$ is total number of strange valence quarks and antiquarks in such hadrons. For instance, $\pi$, $K$ and $\phi$-mesons have, respectively,  $S_\pi=0$,
$S_K=1$ and $S_\phi=2$. 
In contrast to other  versions of HRGM known from the literature (see references in \cite{SFO:13,HRGM:13,Andronic:05}) the present one 
accounts for the hard-core repulsion using different hard-core radii for pions, $R_{\pi}$, kaons, $R_K$,  $\Lambda$-hyperons $R_\Lambda$, 
other mesons, $R_m$, and baryons, $R_b$. 
The best global fit of  111 independent hadronic multiplicities measured in the collision energy range from  $\sqrt{s_{NN}} =2.7$
GeV to $\sqrt{s_{NN}} = 200$ GeV
was found for $R_b$ = 0.355 fm, $R_m$ = 0.4 fm, $R_{\pi}$ = 0.1 fm,  $R_K$ = 0.38 fm   
and $R_\Lambda = 0.11$ fm
with  the quality  $\chi^2/dof \simeq 0.95$ \cite{Sagun}.

Since the main features of the fitting procedure are well documented in papers \cite{SFO:13,HRGM:13}, here we briefly remind 
the basic elements of  HRGM and formulate the new features of present fit. The most important ingredient of the HRGM with 
multicomponent hard-core repulsion is introduced via the matrix of second virial coefficients $b_{ij}$ which  for the hadrons of radii $r_i$ and $r_j$ reads as $b_{ij}=\frac{2\pi}{3}(r_i+r_j)^3$. The hadrons of $i$-th sort are characterized by the spin-isospin degeneracy $g_i$, the mass $m_i$ and the width $\Gamma_i$ which are taken from the particle tables of the thermodynamic code THERMUS \cite{Thermus}.
The HRGM  equation of state defines the set of partial pressures $p_i$ for each hadronic 
component ($p=\sum_i p_i$ is total pressure)
\begin{equation}
\label{EqI}
p_i=T\varphi_i\exp\left[\frac{\mu_i-2\sum_j p_jb_{ji}+\sum_{jl}p_j b_{jl}p_l/p}{T}\right] \,, 
\end{equation}
where $\mu_i=Q^B_i\mu_B+Q^{I3}_i\mu_{I3}+Q^S_i\mu_S$ is the full chemical potential of $i$-th sort of hadrons expressed via the  charges $\{Q^A_i\}$ of $i$-th hadron sort and the corresponding chemical potentials $\{\mu_A\}$.  
In the Boltzmann approximation  the thermal density of $i$-th hadron sort is
\begin{equation}
\label{II}
\varphi_i=\gamma_s^{S_i}g_i \hspace*{-0.1cm}\int\limits_{M_i}^\infty \hspace*{-0.1cm}dm f(m,m_i,\Gamma_i) \hspace*{-0.1cm}
\int \hspace*{-0.1cm}
\frac{d{\bf k}}{(2\pi)^3} \hspace*{0.1cm}e^{- \frac{\sqrt{m^2+{\bf k}^2} }{T} } \,.
\end{equation}
Here $M_i$ is the dominant decay channel threshold and $f$ is the normalized  Breit-Wigner mass attenuation. 
Thermal multiplicities $N_i^{th}=V\frac{\partial p}{\partial\mu_i}$ ($V$ is the effective emission volume) should be 
corrected by the hadron decays after the CFO according to the branching ratios $Br_{l\rightarrow i}$ which define 
the probability of particle $l$ to decay into a particle $i$. Hence  the ratio of full  multiplicities can be cast as
\begin{equation}
\label{III}
R_{ij}\equiv\frac{N^{tot}_i}{N^{tot}_j}=
\frac{p_i+\sum_{l\neq i}p_lBr_{l\rightarrow i}}{p_j+\sum_{l\neq j}p_lBr_{l\rightarrow j}}\,.
\end{equation}
\begin{figure}[t]
%\hspace*{.5cm}
\centerline{
\,\,\,\includegraphics[height=84 mm]{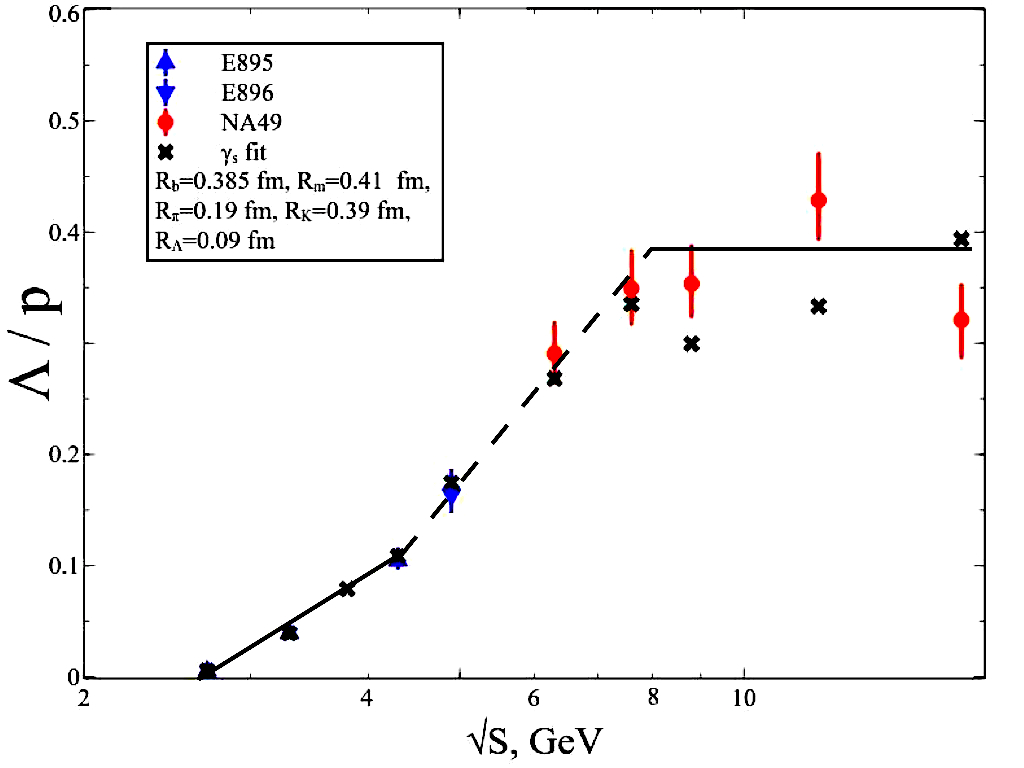} 
}
 \caption{
 The collision energy dependence of the $\Lambda/p$ ratio  obtained within the present HRGM. The lines are given to guide the eye. More explanations tare given in the text. 
}
  \label{fig3}
\end{figure}
This expression was used in 
our analysis of experimental hadron yields since it allows one to exclude $V$ and, hence, to reduce the number of  fitting parameters. 
%%%In addition it turns out that experimental data on ratios usually have smaller errors.  
Here we fit the high quality experimental multiplicities measured at AGS for $\sqrt s_{NN}=2.7, 3.3, 3.8, 4.3, 4.9$ GeV\cite{AGS1,AGS2,AGS2b,AGS3,AGS4,AGS5,AGS6,AGS7,AGS8}, the NA49 data measured at SPS energies  $\sqrt s_{NN}=6.3, 7.6, 8.8, 12.3, 17.3$ GeV  \cite{SPS1,SPS2,SPS3,SPS4,SPS5,SPS6,SPS7,SPS8,SPS9} and the STAR data measured  at 
RHIC energies  $\sqrt s_{NN}=9.2, 62.4, 130, 200$ GeV \cite{RHIC}.
These  experimental data allow us to construct 111 independent ratios measured at 14 values of collision
energies \cite{SFO:13,HRGM:13}. However, from our past experience of successful fitting we learned that at any collision energy it is practically  impossible to  simultaneously reach the mean values of all independent hadron multiplicity ratios and hence the values of error bars are important. On the other hand, we would like to stress that some ratios have essentially smaller error bars than the other ones. Therefore, inclusion of such  dependent ratios into a fit is important because it may affect some  fitting parameters, since their errors are independent on the errors of other ratios. In particular, at AGS and SPS energies the ratio $\Lambda/p$ has rather small error bars (see the upper panel of Fig.\ \ref{fig3}).  One should also remember that an exclusive role of  $\Lambda$-hyperon was demonstrated recently in  \cite{Sagun}, where it was shown that the high quality simultaneous  description of $K^+/\pi^+$,  $\Lambda/\pi^-$ and $\bar \Lambda/\pi^-$ 
ratios requires introduction of an independent hard-core radius for $\Lambda$ hyperons.  
Therefore, to study the role of the  $\Lambda/p$  ratio  on the total fit quality we 
included it into our data sample and obtained 121 hadron multiplicity  ratios to analyze.

The local fitting parameters for each collision  energy ($T$, $\mu_B$, $\mu_{I3}$ and $\gamma_s$) and the global ones, i.e. 
the values of  hard-core radii $R_b$ = 0.385 fm, $R_m$ = 0.41 fm, $R_{\pi}$ = 0.19 fm,  $R_K$ = 0.39 fm   
and $R_\Lambda = 0.09$ fm, were determined by minimizing the quantity
\begin{equation}\label{EqIV}
\chi^2=\sum_{p=1}^{121}\frac{(R_p^{theor}-R_p^{exp})^2}{\sigma_p^2},
\end{equation}
where the summation is carried out over all data points.
Here $\sigma_p$ is an experimental error of ratio $R_p^{exp}$. The  minimization of (\ref{EqIV}) showed that compared to previous results \cite{SFO:13,HRGM:13,Sagun} the values of  local fitting parameters
$T$, $\mu_B$, $\mu_{I3}$ and $\gamma_s$ are practically the same (i.e. within the errors bars they simply coincide), while the values of hard-core radii are changed.  The present fit corresponds to a slight increase of  $R_b$, $R_m$ and  $R_K$. This makes their values practically the same, while $R_\Lambda $ demonstrate a slight decrease. The largest increase is experienced by the pion hard-core radius $R_{\pi}$, which almost doubles. The  achieved fit quality $\chi^2/dof = 63.98/65 \simeq 0.98$ is about 3\% larger than the one found in \cite{Sagun}. The results of $\Lambda/p$ fit are shown in the upper panel of Fig.\ \ref{fig3}. Compared to the $\Lambda/p$ ratio which was found   in \cite{Bugaev:SA2}  (see Fig.\ 14 in \cite{Bugaev:SA2}) without including it  into a fitting procedure, the results of the upper panel of Fig.\ \ref{fig3} demonstrate a sizable improvement of  the $\Lambda/p$ ratio for $\sqrt{s_{NN}} =2.7-7.9$ GeV, while at  higher collision energies the description of this ratio slightly worsens. The quality of description of other ratios remains the same as in \cite{Sagun}.  Also we checked that the sharp peak of the trace anomaly at CFO (see the upper panel of  Fig.\ \ref{fig1}) is not affected by the change of  pion hard-core radius. It turs out that  inclusion of  the $\Lambda/p$ ratio into a fit is important because it improves the description of this ration and, hence, it provides a higher confidence in our predictions for NICA and FAIR experiments which are related to the $\Lambda/p$ ratio. 

\begin{figure}[t]
\centerline{\,\,\includegraphics[height=84mm]{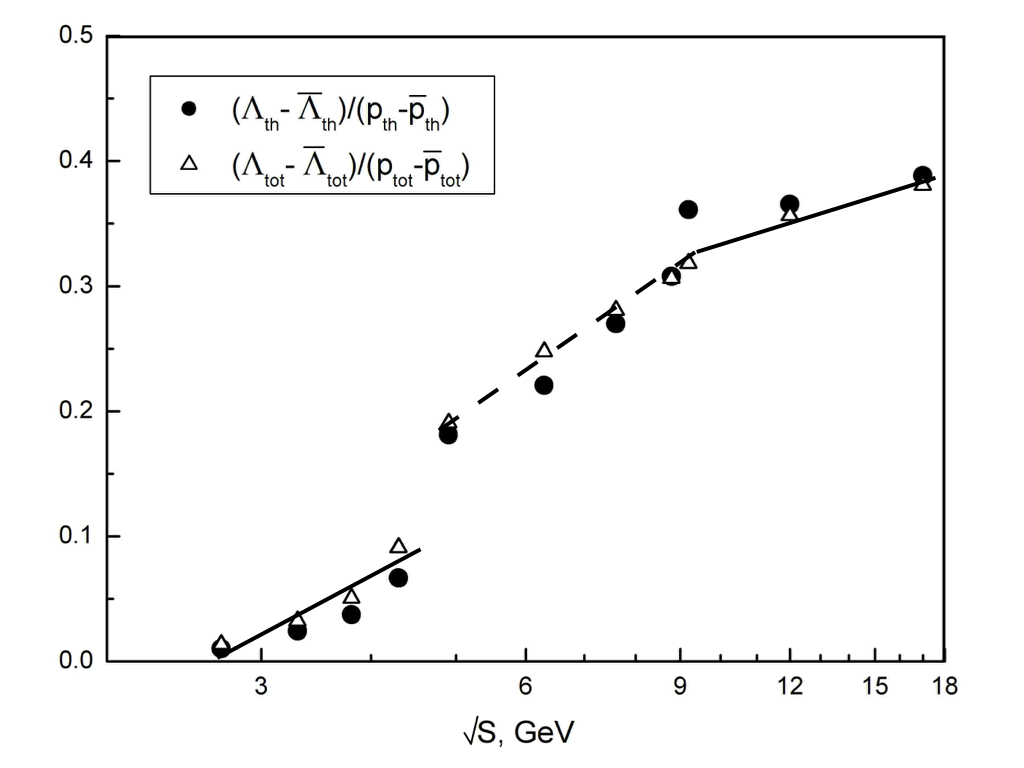}}
\vspace*{-1.5mm}
\centerline{\includegraphics[height=84 mm]{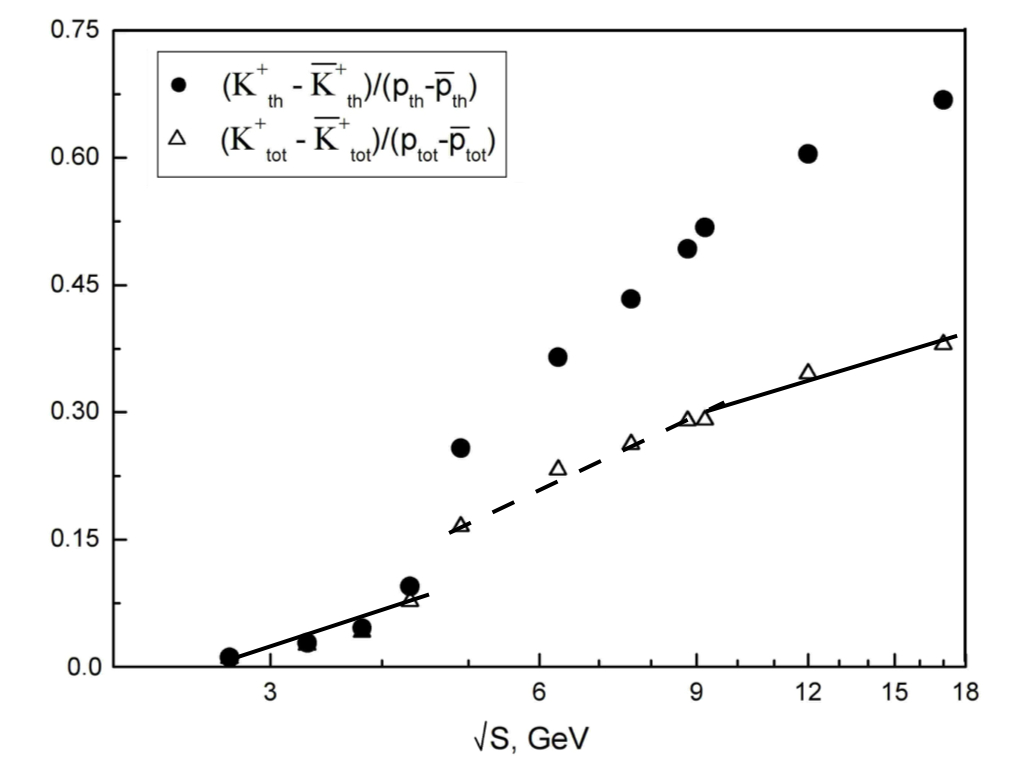}}
 \caption{Predictions for the collision energy dependence of  $\frac{\Delta \Lambda}{\Delta\, p}$ (upper panel) and  $\frac{\Delta K^+}{\Delta \,p}$ (lower panel)  ratios. The triangles depict the total multiplicities, while the circles correspond to the thermal multiplicities. The lines are given to guide the eye. 
}
  \label{fig4}
\end{figure}

Fig.\ \ref{fig3} clearly shows that there are three regimes in the
energy dependence of  the $\Lambda/p$
ratio: at  $\sqrt{s_{NN}}=4.3$ GeV the slope of this ratio clearly increases, while above   
$\sqrt{s_{NN}}=7.6$ GeV it nearly saturates.  
The change of slope at  $\sqrt{s_{NN}}=4.3$ GeV is
very similar to the prediction of Ref. \cite{Rafelski:82} that the mixed phase formation can be 
identified by a rapid increase in the number of strange quarks per light quarks.  
This  ratio is a  convenient indicator because at low collision energies $\Lambda$ hyperons
are generated in collisions of nucleons. As it is seen from Fig.\ \ref{fig3}, this mechanism works up to  
$\sqrt{s_{NN}}=4.3$ GeV,
while an appearance of the mixed phase should lead to an  increase 
{in the} number of strange quarks and antiquarks 
due to {the} annihilation of light quark-antiquark {and gluon pairs}.
Note that this simple  picture is in line with the result that the mixed phase can be reached at  $\sqrt{s_{NN}}=4.3$ GeV,
while QGP is formed at $\sqrt{s_{NN}} \ge 4.9$ GeV.  The dramatic change  of the experimental  $\Lambda/p$
ratio slope at $\sqrt{s_{NN}} > 7.6$ GeV which is seen in  Fig.\ \ref{fig3} can be an evidence for the second 
phase transformation, which we discussed in the preceding section.  

From   Fig.\ \ref{fig4} one can see even more dramatic changes in the collision energy dependence  of  two new ratios $\frac{\Delta \Lambda}{\Delta\, p} = \frac{\Lambda - \bar \Lambda}{p - \bar p}$ and 
$\frac{\Delta K^+}{\Delta \,p} = \frac{K^+ - \bar K^+}{p~ -~ \bar p}$.
Both quantities demonstrate 
a strong jump right in the collision energy region which is associated with the mixed phase formation, i.e. for  $\sqrt{s_{NN}} =4.3-4.9$ GeV. Also in Fig.\ \ref{fig4}  one can clearly see a change of  the $\frac{\Delta \Lambda}{\Delta\, p}$  and
$\frac{\Delta K^+}{\Delta \,p}$ slopes  at $\sqrt{s_{NN}} = 9.2$ GeV. 
 Our hypothesis is that the collision energy dependence  of  the  $\frac{\Delta \Lambda}{\Delta\, p}$  and
$\frac{\Delta K^+}{\Delta \,p}$ ratios  is an indicator of two phase transformations. Since the observed jump of these ratios  is located in  the collision energy region of the mixed phase formation (i.e. a first order phase transition), then a change of  their  slope at $\sqrt{s_{NN}} = 9.2$ GeV  can be naturally associated with a weak first order or a second order phase transition.  Note that such a hypothesis is well supported by  the results of the meta-analysis \cite{Metaanalisys} which we  summarized above. 

In contrast to other multiplicity ratios  these ratios have a clear physical meaning of  strange charge of $\Lambda$ hyperons (of $K^+ $ mesons) per baryon charge of protons.  Also we would like to stress that at low collision energies the antibaryons are  absent and, hence, $\frac{\Delta \Lambda}{\Delta\, p} \simeq \Lambda/p$. This can be  easily seen from a comparison of    Fig.\ \ref{fig3} with the upper panel of Fig.\  \ref{fig4}.  As one can see from the upper panel of  Fig.\ \ref{fig4} the behavior of
thermal multiplicities of the $\frac{\Delta \Lambda}{\Delta\, p}$ ratio is very similar to the one of total multiplicities at CFO
and, therefore, such a ratio maybe used to directly access the moment of CFO by other models which, in contrast to HRGM, do not account for the resonance decays.

\section{Summary}
Here we briefly describe  the peculiar irregularities at chemical freeze-out which, as we argue, can be used as the signals of the mixed phase/QGP formation. Two entirely different approaches presented here  give the evidence that there is a possibility to experimentally detect two phase transitions in QCD. First of them is the first order phase transition to the quark-gluon-hadron mixed  phase
which, as we expect, can be formed at collision energies  $\sqrt{s_{NN}}=4.3-4.9$ GeV.  The nature of the second phase transition is not fully understood now, but we are arguing that this can be either the weak first order  or the second order phase transition, which can be probed at collision energies  $\sqrt{s_{NN}}=10-13.5$ GeV.  In the former case the initial states formed at the beginning of nuclear-nuclear collision  leave  QGP   and return back to the mixed phase at these collision energies. In the latter 
case the initial states formed at the beginning of nuclear-nuclear collision cross the curve of the second order phase transition which maybe either the (partial) restoration of chiral symmetry or a  transition between the quarkyonic matter and QGP \cite{QYON}. Note that the irregularities/signals found at these collision energies can be studied at NICA and FAIR. 

In this work we also present  the results of an advanced fit of 121 hadron multiplicity ratios measured at AGS, SPS and RHIC. The HRGM with multicomponent repulsion allowed us to get a nearly perfect  description of the $\Lambda/p$ ratio for the collision energies $\sqrt{s_{NN}} =2.7-7.9$ GeV and to keep the high quality of  description of  other hadronic multiplicity ratios with the total quality 
$\chi^2/dof \simeq 0.98$. Such a high fit quality gives us a high confidence that the suggested ratios $\frac{\Lambda - \bar \Lambda}{p - \bar p}$ and $\frac{K^+ - \bar K^+}{p~ -~ \bar p}$ can serve as  safe indicators of discussed phase transitions. Hopefully,  such ratios can be used at NICA and FAIR  to locate the  threshold energies of these transitions.

\vskip2.2mm

{\bf Acknowledgements.} The authors are thankful to D. B. Blaschke, T. Galayuk, R.A. Lacey, I. N. Mishustin, D. H. Rischke, K. Redlich, L. M. Satarov  and K. Urbanowski for discussions and valuable comments. K.A.B., V.V.S., A.I.I, D.R.O. and G.M.Z. acknowledge the  partial support of the program ``On perspective fundamental research in high-energy and nuclear physics'' launched by the Section of Nuclear Physics of NAS of Ukraine.

%%%\vspace*{2.2cm}
%%%KAB

\end{document}